\documentclass[aps,twocolumn,float,prd,psfig]{revtex4}

\usepackage{color}

\input epsf
\usepackage{graphicx}
\newcommand{\beq}{\begin{equation}}
\newcommand{\beqa}{\begin{eqnarray}}
		  \newcommand{\eeq}{\end{equation}}
\newcommand{\eeqa}{\end{eqnarray}}

\newcommand{\lsim}{\lesssim}
\newcommand{\gsim}{\gtrsim}

\newcommand{\vect}[1]{\mbox{\boldmath${#1}$}}
\newcommand{\lmk}{\left(}
\newcommand{\rmk}{\right)}

\newcommand{\lkk}{\left[}
\newcommand{\rkk}{\right]}
\newcommand{\lla}{\left\langle}

\newcommand{\rra}{\right\rangle}

\newcommand{\vex}{{\vect x}}

\newcommand{\veh}{{\vect h}}
\newcommand{\vue}{\hat{\vect e}}
\newcommand{\vel}{\vect l}
\newcommand{\veD}{\vect D}

\newcommand{\ven}{\vect n}
\newcommand{\vem}{\vect m}

\newcommand{\veu}{{\vect u}}
\newcommand{\vev}{{\vect v}}

\newcommand{\ve}{{\vect e}}

\newcommand{\cp}{\rm {c_{\alpha}}}
\newcommand{\cq}{\rm {c_{\beta}}}
\newcommand{\cpq}{\rm {c_{\alpha\beta}}}

\begin{document}
\title{Gravitational Wave Background Search by Correlating Multiple Triangular Detectors \\
in the mHz Band
 } 
%
%
%
\author{Naoki Seto }
\affiliation{Department of Physics, Kyoto University, 
Kyoto 606-8502, Japan
}
\date{\today}
%
%
%
%
%
%
\begin{abstract}
 With the recent strong  developments of TianQin and  Taiji, we now have an increasing chance to make a correlation analysis in the mHz band by operating them together with LISA.  Assuming two  LISA-like triangular detectors at general geometrical configurations,  we develop a simple formulation to evaluate  the network sensitivity to an  isotropic gravitational wave background. In our formulation, we fully use the symmetry of data channels within each triangular detector and provide tractable expressions without directly employing cumbersome detector tensors.   We concretely evaluate the expected network sensitivities for various potential detector combinations, including the LISA-TianQin pair.

\end{abstract}

\maketitle

\section{Introduction}

A cosmological gravitational wave background is  an important observational target for studying our universe \cite{Romano:2016dpx,Caprini:2018mtu}. It is expected to have  a nearly isotropic intensity  pattern. 
We can statistically amplify the background signal by correlating noise independent detectors \cite{Christensen:1992wi,Flanagan:1993ix,Allen:1997ad} (see also \cite{Hellings:1983fr}). 
  Around 10-1000 Hz,  the observational constrains have been significantly improved  with the suppression of instrumental noises of ground based detectors. The latest result by the LIGO-Virgo collaboration  is $\Omega_{GW}<6.0\times 10^{-8}$ \cite{LIGOScientific:2019vic}.   On another front, quite recently, the nano-Hertz band is  fueled by a report from NANOGrav \cite{Arzoumanian:2020vkk}.

LISA has been the leading project for gravitational wave observation around $10^{-4}$-0.1Hz \cite{lisa0,lisa}.  It is scheduled to be launched around 2035. From its triangular detector, we can take three noise independent data channels \cite{Prince:2002hp}. 
Unfortunately, due to symmetric cancellations,  the correlations between the noise independent data channels are known to be  insensitive to the monopole pattern of a background \cite{Romano:2016dpx} (see also \cite{Tinto:2001ii,Hogan:2001jn,Adams:2010vc} for a different approach with Sagnac-type data streams). But, recently, the situations have been changed rapidly with the strong developments of two Chinese projects; TianQin \cite{Luo:2015ght} and Taiji \cite{taiji} (see Fig.1 for their noise spectra). In this context, the author studied the prospects specifically for the LISA-Taiji network \cite{Seto:2020zxw} (see also \cite{Cornish:2001bb,1821247,Ruan:2019tje}).  He pointed out two important symmetries inherent to the network.  The first one is the geometric symmetry about the relative configuration of  LISA and Taiji. We can take a virtual sphere that simultaneously contacts with the two detector planes. The second symmetry is about the combinations of data channels within each triangular detector (related to the insensitivity to the monopole pattern mentioned above). It was shown that, as a result of the two symmetries,  the LISA-Taiji network  allows us to make a simple parity decomposition for an isotropic  gravitational wave background  \cite{Seto:2020zxw}.

\begin{figure}[t]
 \includegraphics[width=8.cm,angle=0,clip]{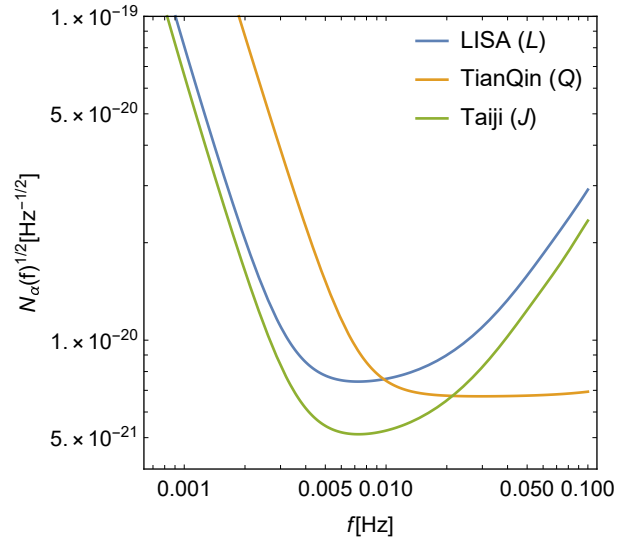}
 \caption{The instrumental noise spectra for the planned triangular detectors.    Each curve is  given for a single ($A$ and $E$) data channel without angular averages. 
} 
\end{figure}


In this paper,  we study general network configurations for two LISA-like triangular detectors. Now, in contrast to the LISA-Taiji network, the first geometrical symmetry mentioned above  is no longer available.  But we can still exploit the second symmetry about the internal data channels. Indeed, taking the advantage of  the second one,  we only need  three angular parameters for characterizing the homothetic structure of the network geometry.  This enables us to develop a simple formulation for evaluating the network sensitivities to an isotropic background.  In this formulation, without directly handling cumbersome detector tensors,  we  calculate  some polynomials given by the cosines of the three angular parameters.  
Using our  formulation, we concretely evaluate the sensitivities for networks formed by pairing LISA, TianQin, Taiji and their variations. 

This paper is organized as follows.  In Sec.II, we review the basic aspects of the correlation analysis. In Sec.III, we study the networks composed by two LISA-like triangular detectors and develop a new formulation, fully using the symmetry of their data combinations. In Sec.IV, we explain how our analytical expressions are simplified for networks with high geometrical symmetries. In Sec.V, we apply our formulation to the mHz-band networks formed by LISA, TianQin, Taiji and their variations. In Sec.VI, we numerically evaluate the signal-to-noise ratios for  potential networks.

\section{correlation analysis}
In this section, we make a brief discussion on the correlation analysis for detecting an isotropic gravitational wave background (see \cite{Romano:2016dpx,Flanagan:1993ix,Allen:1997ad} for detail). 
\subsection{background signal}
 We can expand the metric perturbation at a position $\vex$ induced by a gravitational wave background as  
\beq
\veh(f,\vex)=\sum_{P={+,\times}} \int d\ven\, h(f\ven,P) \ve_p \,e^{-2\pi i f \ven\cdot \vex/c  }.\label{expa}
\eeq
Here the unit vector $\ven$ shows the propagation direction of a wave, and the factor $ h(f\ven,P)$ is the  mode coefficient.  The polarization tensors $\ve_P, (P=+,\times )$ are given by the orthonormal transverse vectors $\vel_1$ and $\vel_2$ as
\beqa
\ve_+&=&\vel_1\otimes  \vel_1-\vel_2\otimes  \vel_2\\
\ve_\times&=&\vel_1\otimes  \vel_2+\vel_2\otimes  \vel_1.
\eeqa
In this paper, as  Eq.(\ref{expa}), we mainly use quantities in  the Fourier space rather than the time space. 

We assume that the gravitational wave background is isotropic, stationary and randomly polarized. Then, the covariance of the mode coefficients is written by
\beq
\lla  h(f\ven,P)  h(f\rq{}\ven\rq{},P\rq{})^*   \rra= \delta (f\ven-f\rq{}\ven\rq{})\delta_{P,P\rq{}}S_h(f). \label{sh}
\eeq
In general relativity, the strain  spectrum $S_h(f)$ is given by the normalized energy density $\Omega_{GW}(f)$ as 
\beq
S_h(f)=\frac{3H_0^2}{32\pi^3} \Omega_{GW}(f)
\eeq
with the Hubble parameter $H_0$ that is fixed at $70{\rm km^{1} s^{-1} Mpc^{-1}}$ below.

\subsection{cross correlation}

Next we consider an equal-arm  L-shaped interferometer $a$ at a position  $\vex_a$, and discuss its response $h_a(f)$ to the background. 
In the low frequency approximation, the response is characterized by the detector tensor 
\beq
\veD_{a}=( \veu\otimes\veu-\vev\otimes\vev)/2
\eeq
with the unit vectors $\veu$ and $\vev$ representing the two arm directions  ($\veu\cdot\vev=0$). 
Then, the interferometric response to the background is given by  
\beq
h_a(f)=\sum_{P={+,\times}}\int d\ven\, h(f\ven,P)  \,e^{-2\pi i f \ven\cdot \vex_a/c  } F_{a}^P(\ven) \label{ha}
\eeq
with the beam pattern function $F_{a}^P(\ven)=(\veD_a:\ve_P)$.
In reality,  the data stream $s_a(f)$  of the detector $a$ contains not only the signal $h_a(f)$ but also various noises $n_a(f)$.  Therefore, we put
\beq
s_a(f)=h_a(f)+n_a(f).
\eeq
The noise spectrum $N_a(f)$ is defined by 
\beq
\lla n_a(f) n_a(f\rq{})^*\rra=N_a(f)\delta(f-f\rq{}) .
\eeq

The correlation analysis is an efficient approach to detect a weak background signal dominated  by the noise $|h_a(f)|\ll |n_a(f)|$. The expectation value of the correlation product of two noise-independent detectors $a$ and $b$ is given by 
\beqa
\lla s_a(f) s_b(f\rq{})^*\rra&=&
\lla h_a(f) h_b(f\rq{})^*\rra\\
&=&\frac{8\pi}5\delta(f-f\rq{}) \gamma_{ab}(f)S_h(f).\label{ss}
\eeqa
Here $\gamma_{ab}(f)$ is the overlap reduction function. From Eqs.(\ref{sh}) and (\ref{ha}), we have \cite{Flanagan:1993ix,Allen:1997ad}
\beq
\gamma_{ab}(f)=\frac5{8\pi}\sum_{P=+,\times}\int d\ven F_{a}^P(\ven) F_{b}^P(\ven)e^{2\pi if\ven\cdot(\vex_b-\vex_a)}.\label{orf}
\eeq
Considering the diagonal structure in Eq.(\ref{ss}),  we take the product of the data at same frequencies $f=f\rq{}$.  The fluctuation around the expectation value (\ref{ss}) is assumed to be dominated by the noises $n_a(f)n_b(f)^*$.
In order to statistically  suppress the noises, we take a summation of a large number of Fourier bins with the frequency interval $T_{\rm obs}^{-1}$ ($T_{\rm obs}$: observation period). If we take the integration range $f\in [f_{\rm min},f_{\rm max}]$, the signal-to-noise ratio is given by \cite{Flanagan:1993ix,Allen:1997ad}
\beqa
SNR^2&=&2\lmk \frac{3H_0^2}{10\pi^2} \rmk^2 T_{\rm obs}\nonumber\\
& & \times   \lkk 	\int_{f_{\rm min}}^{f_{\rm max}}  df \frac{\Omega_{\rm GW}(f)^2}{f^6}  \frac{ \gamma_{ab}(f)^2}{N_{a}(f)N_{b}(f)}  \rkk.  
\eeqa
When we have more than one pair of noise-independent interferometers, the element $\gamma_{ab}(f)^2/(N_a(f)N_b(f))$ should be replaced by the summation of the corresponding combinations.

\subsection{overlap reduction function}
In Eq.(\ref{orf}), we defined  the  overlap reduction function $\gamma_{ab}(f)$. 
 Here  we provide its tensorial expression that will be extensively used in the next section. 

Let us consider two L-shaped interferometers $a$ and $b$ separated at
\beq
\vex_b-\vex_a=d \vem 
\eeq
with the distance $d=|\vex_a-\vex_b|$ and the normalization $|\vem|=1$. 
Then, in the low frequency approximation,  the overlap reduction can be formally expressed as 
\beq
\gamma_{ab}(f)=\Gamma_{ijkl}D_{a,ij} D_{b,kl} \label{gab}
\eeq
with the two detector tensors $D_{a,ij}$ and $D_{b,ij}$ \cite{Flanagan:1993ix,Allen:1997ad}. 
The tensor $\Gamma_{ijkl}$ is written by the Kronecker\rq{}s delta and the unit vector $\vem$ as
\beqa
\Gamma_{ijkl}(y)&=&b_{0}(y) \delta_{ik}\delta_{jk}+b_1(y)\delta_{ik}m_jm_l\nonumber \\
& & +b_2 (y) m_im_jm_km_l.
\eeqa
The coefficients $b_0$, $b_1$ and $b_2$ are given by the spherical  Bessel functions $j_l=j_l(y)$ with the argument
$y=2\pi f d/c$ as
\beqa
b_0(y)&=&2\lmk j_0-\frac{10}7 j_2+\frac1{14}j_4   \rmk \label{b0} \\
b_1(y)&=&4\lmk \frac{15}7 j_2-\frac5{14}j_4   \rmk \\
b_2(y)&=&\frac52j_4.\label{b2}
\eeqa

\begin{figure}
 \includegraphics[width=8.cm,angle=270,clip]{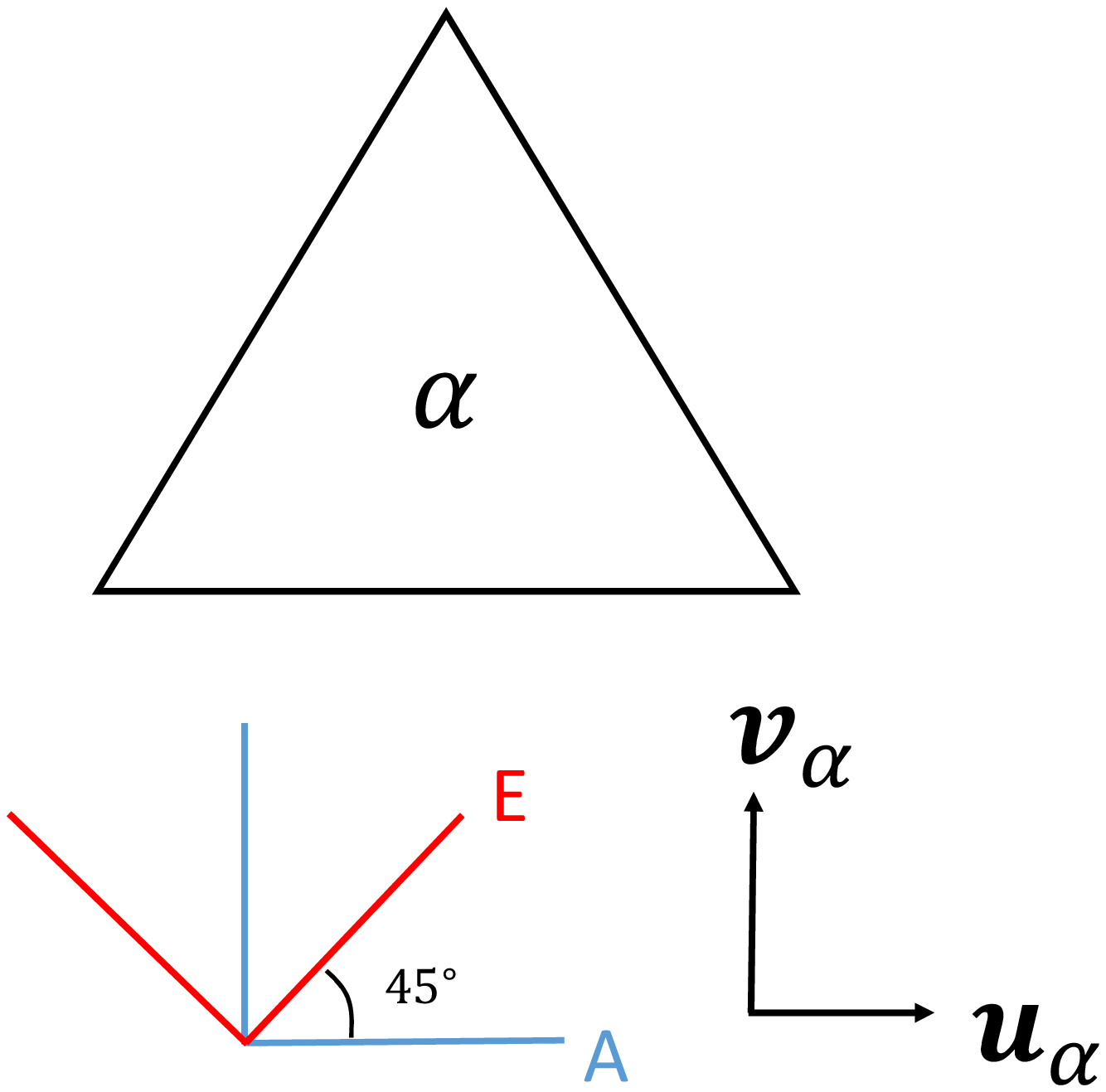}
 \caption{ The geometry of the two effective L-shaped interferometers for the $A$ and $E$ channels generated from a single triangle unit $\alpha$. Their detector tensors $\veD_A$ and $\veD_E$ are characterized by the unit vectors $\veu$ and $\vev$ as in Eqs.(\ref{da}) and (\ref{de}). 
 These detector tensors are attached to the triangle that spins as a result of differential orbital motions.} 
\end{figure}


\section{triangular detectors}
In this section, based on the results presented in the previous section, we study the correlation analysis specifically for two LISA-like triangular interferometers.  
 We use the labels  $\alpha$ and $\beta$ to specify the triangular units.

\subsection{A and E channels}

First, we discuss data streams available from a single LISA-like triangular unit $\alpha$.  We can, in principle, make three interferometers symmetrically at the three vertexes. However, these data streams have correlated noises. 
Using the symmetry of the system, we can easily compose three data channels $A, E$ and $T$ with no correlated noises \cite{Prince:2002hp}.
The $T$ channel has a negligible strain sensitivity at the frequency regime relevant for our study, and we thus deal with the $A$ and $E$ channels below.  They can be effectively regarded as two L-shaped interferometers with the offset angle $45^\circ$ (see Fig.2).

Using the orthonormal unit vectors $(\veu_\alpha,\vev_\alpha)$ attached to the triangle $\alpha$, the detector tensor for the $A$ channel is given by 
\beq
\veD_{A}=( \veu_\alpha\otimes\veu_\alpha-\vev_\alpha\otimes\vev_\alpha)/2.\label{da}
\eeq
The arm directions of the $E$ channel are given by $(\veu_\alpha\pm \vev_\alpha)/\sqrt2$, and we obtain
\beq
\veD_{E}= (\veu_\alpha\otimes\vev_\alpha+\vev_\alpha\otimes\veu_\alpha)/2.\label{de}
\eeq
With the low frequency approximation, we can readily confirm $\gamma_{AE}\propto (\veD_A:\veD_E)=0$. 
In  fact, due to the symmetry of the system, the $AE$ correlation is  insensitive  to an isotropic background even without the low frequency approximation.

For the noise spectra of the $A$ and $E$ channels, we  have 
\beq
\lla n_A(f) n_A(f\rq{})^*\rra=\lla n_E(f) n_E(f\rq{})^*\rra=N_\alpha(f)\delta(f-f\rq{}) \label{noi1}
\eeq
with no correlation
\beq
\lla n_{A}(f) n_{E}(f\rq{})^*\rra=0. \label{noi2}
\eeq

\subsection{total SNR}

Next we study correlation analysis with two triangular detectors $\alpha$ and $\beta$. Similar to the triangle $\alpha$ discussed in the previous section, we take two noise independent data streams $(A\rq{},E\rq{})$ from  the second triangle $\beta$, and put their noise spectra by $N_\beta(f)$.   Throughout  this paper we assume that two triangles have no noise correlation. 

In total, we have four data pairs $AA\rq{},AE\rq{}, EA\rq{}$ and $EE\rq{}$.  The total signal-to-noise ratio is given by
\beq
SNR_{\alpha\beta}^2=\lmk \frac{3H_0^2}{10\pi^2} \rmk^2 T_{\rm obs}  \lkk 2\int_{f_{\rm min}}^{f_{\rm max}}  df \frac{\Omega_{\rm GW}(f)^2 Y_{\alpha\beta}(f)}{f^6N_{\alpha}(f)N_{\beta}(f)}   \rkk   \label{snr0}
\eeq
where we defined the function  $Y_{\alpha\beta}(f)$ by 
\beq
Y_{\alpha\beta}(f)\equiv \gamma_{AA\rq{}}^2+\gamma_{AE\rq{}}^2+\gamma_{EA\rq{}}^2+\gamma_{EE\rq{}}^2,
\eeq
using the four overlap reduction functions.  This function  $Y_{\alpha\beta}(f)$ plays a central role in the rest of this paper.  We call it the total response function.

Using Eqs.(\ref{gab})-(\ref{b2}), we can formally expand 
\beq
Y_{\alpha\beta}=~\sum_{i=0}^2\sum_{j=0}^2 b_i(y) b_j(y) X_{ij} . \label{yab}
\eeq
Here $X_{ij}(=X_{ij})$ are given by the detector tensors $D_{a,ij}$ and the unit directional vector $m_i$. For example, we have 
\beq
X_{00}=\sum_{a}^{AE}\sum_{b}^{A\rq{}E\rq{}}(\delta_{ik}\delta_{jl} D_{a,ij} D_{b,kl})  (\delta_{tr}\delta_{us} D_{a,rs} D_{b,tu}), \label{x00}
\eeq
\beq
X_{01}=\sum_{a}^{AE}\sum_{b}^{A\rq{}E\rq{}}(\delta_{ik}\delta_{jl} D_{a,ij} D_{b,kl})  (\delta_{tr}m_u m_s D_{a,rs} D_{b,tu}).\label{x01}
\eeq

In Eq.(\ref{yab}), the factors $b_i(y)$ ($i=0,1$ and 2) are given by the spherical Bessel functions, and  closely related to the wave effects with the argument $y=2\pi fd/c$.  In contrast,  the factors $X_{ij}$ contain tensorial information, and are determined by the homothetic structure  of the network.  Hereafter, we call  $b_i(y)$ by the wave factors  and  $X_{ij}$ by the tensorial factors.

\begin{figure}
 \includegraphics[width=3.5cm,angle=270,clip]{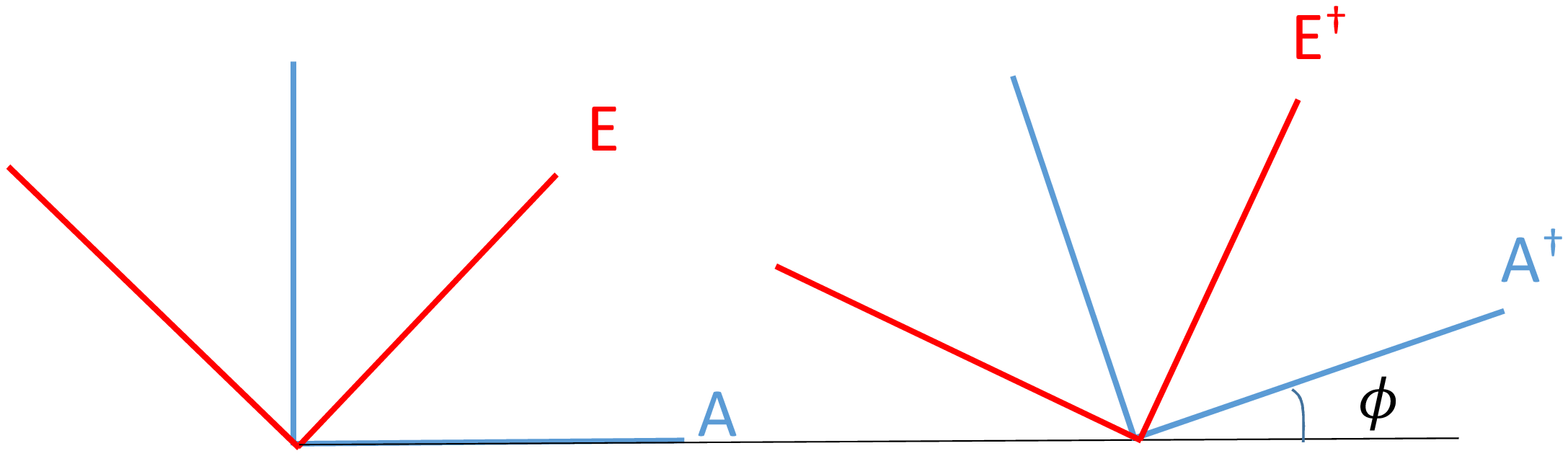}
 \caption{The virtual rotation of the detector tensors induced by the linear combination (\ref{aer}) of  the two data channels $(A,E)$.
} 
\end{figure}


\subsection{virtual rotation}\label{vr}

In Sec.III A, we explained the noise independent data channels $(A,E)$ available from a single triangular detector.  In fact, 
 these two channels have an interesting property that becomes  useful at correlating multiple triangular detectors (discussed in the next subsection).

To begin with, we consider the new data channels $(A^\dagger,E^\dagger)$  given as the linear combinations of the original ones $(A,E)$  \cite{Seto:2004ji,Seto:2020zxw,1821247}
\beq
\left(
    \begin{array}{c}
      {s_{A^\dagger}} \\ 
       s_{E^\dagger}
    \end{array}
  \right)
= \left(
    \begin{array}{cc}
      \cos 2\phi& \sin 2\phi \\ 
       -\sin2\phi  & \cos2\phi
    \end{array}
  \right) \left(
    \begin{array}{c}
      s_{A} \\ 
       s_{E}
    \end{array}
  \right) \label{ae}.
\eeq

Correspondingly, the detector tensors $(\veD_{A^\dagger}, \veD_{E^\dagger})$  of the new channels are  given by
\beq
\left(
    \begin{array}{c}
     \veD_{A^\dagger} \\ 
      \veD_{E^\dagger}
    \end{array}
  \right)
= \left(
    \begin{array}{cc}
      \cos 2\phi& \sin 2\phi \\ 
       -\sin2\phi  & \cos2\phi
    \end{array}
  \right) \left(
    \begin{array}{c}
     \veD_{A}\\ 
      \veD_{E}
    \end{array}
  \right) \label{aer}.
\eeq
This might look a merely formal transformation.
But, after simple tensorial calculations, we  can    confirm that $(\veD_{A^\dagger}, \veD_{E^\dagger})$ are identical to the tensors obtained by rotating the original detector tensors   $(\veD_{A}, \veD_{E})$ with the angle $\phi$ (see Fig.3). The factor 2 is due to the spin-2 nature of the detector tensors.  Therefore, by taking the linear combinations (\ref{ae}), we  can virtually rotate the two L-shaped detectors. 

Using the relations (\ref{noi1}) and (\ref{noi2}), it is also straightforward to show the noise properties of the new channels $(A^\dagger,E^\dagger)$ as 
\beq
\lla n_{A^\dagger}(f) n_{A^\dagger}(f\rq{})^*\rra=\lla n_{E^\dagger}(f) n_{E^\dagger}(f\rq{})^*\rra=N_\alpha(f)\delta(f-f\rq{}) \label{noi3}
\eeq
and 
\beq
\lla n_{A^\dagger}(f) n_{E^\dagger}(f\rq{})^*\rra=0.\label{noi4}
\eeq
In fact, the three data channel $A,E$ and $T$ are eigenvectors of the instrumental noise matrix. The $A$ and $E$ channels have an identical eigenvalue and are orthonormal \cite{Prince:2002hp}.  As a result, we have freedom to readjust them with a two-dimensional rotation matrix.

\subsection{symmetry of the total response function}
  For the triangle $\alpha$, considering the arguments in the previous subsection, the virtually rotated  channels $(A^\dagger,E^\dagger)$ are equivalent to the original ones $(A,E)$.

Here we again consider the correlation analysis with the  two triangles $\alpha$ and $\beta$. In contrast to Sec.IV B, we use the virtually  rotated channels  $(A^\dagger,E^\dagger)$ for $\alpha$ and the original ones $(A\rq{},E\rq{})$ for $\beta$.
 From the correspondence of the polarization tensors (\ref{aer}), the overlap reduction function for the $A^\dagger A\rq{}$ pair is given by 
\beq
\gamma_{A^\dagger A\rq{}} = \gamma_{AA\rq{}}\cos2\phi +\gamma_{EA\rq{}}\sin2\phi .
\eeq
Similarly, we have 
\beqa
\gamma_{A^\dagger E\rq{}} &=& \gamma_{AE\rq{}}\cos2\phi +\gamma_{EE\rq{}}\sin2\phi ,\\
\gamma_{E^\dagger A\rq{}} &=& -\gamma_{AA\rq{}}\sin2\phi +\gamma_{EA\rq{}}\cos2\phi ,\\
\gamma_{E^\dagger E\rq{}} &=& -\gamma_{AE\rq{}}\sin2\phi +\gamma_{EE\rq{}}\cos2\phi.
\eeqa
Then we obtain 
\beqa
& & \gamma_{A^\dagger A\rq{}}^2+\gamma_{A^\dagger E\rq{}}^2+\gamma_{E^\dagger A\rq{}}^2+\gamma_{E^\dagger E\rq{}}^2\\
&=&\gamma_{AA\rq{}}^2+\gamma_{AE\rq{}}^2+\gamma_{EA\rq{}}^2 +\gamma_{EE\rq{}}^2\nonumber. \\
&=&Y_{\alpha\beta}
\eeqa

This geometrically means that, with respect to the  the total response function $Y_{\alpha\beta}$, we do not have a preferred rotation angle $\phi$ for the two L-shaped interferometers associated with  $\alpha$. In fact, this rotational invariance holds not only for the function $Y_{\alpha\beta}$ but also for each tensorial factor $X_{ij}$.  We can easily understand this from the quadratic dependence on the detector tensors $D_{a,ij}$ as shown in Eqs.(\ref{x00}) and (\ref{x01}).   From the symmetry of the system, we can make the same  arguments for the second triangle  $\beta$.

\subsection{expressions with three cosines}

The tensorial factors $X_{ij}$ are given by the unit directional vector $\vem$ and the four detector tensors $\veD_{a,ij}$ ($a=A,E,A\rq{},E\rq{})$ generated from the two triangles $\alpha$ and $\beta$. We need rather complicated manipulations for evaluating the tensorial factors $X_{ij}$ with the expressions such as Eqs.(\ref{x00}) and (\ref{x01}).  In this subsection, we examine how we can simplify the tensorial factors by using the rotational symmetry discussed in the previous subsection.

\begin{figure}
 \includegraphics[width=8.cm,angle=270,clip]{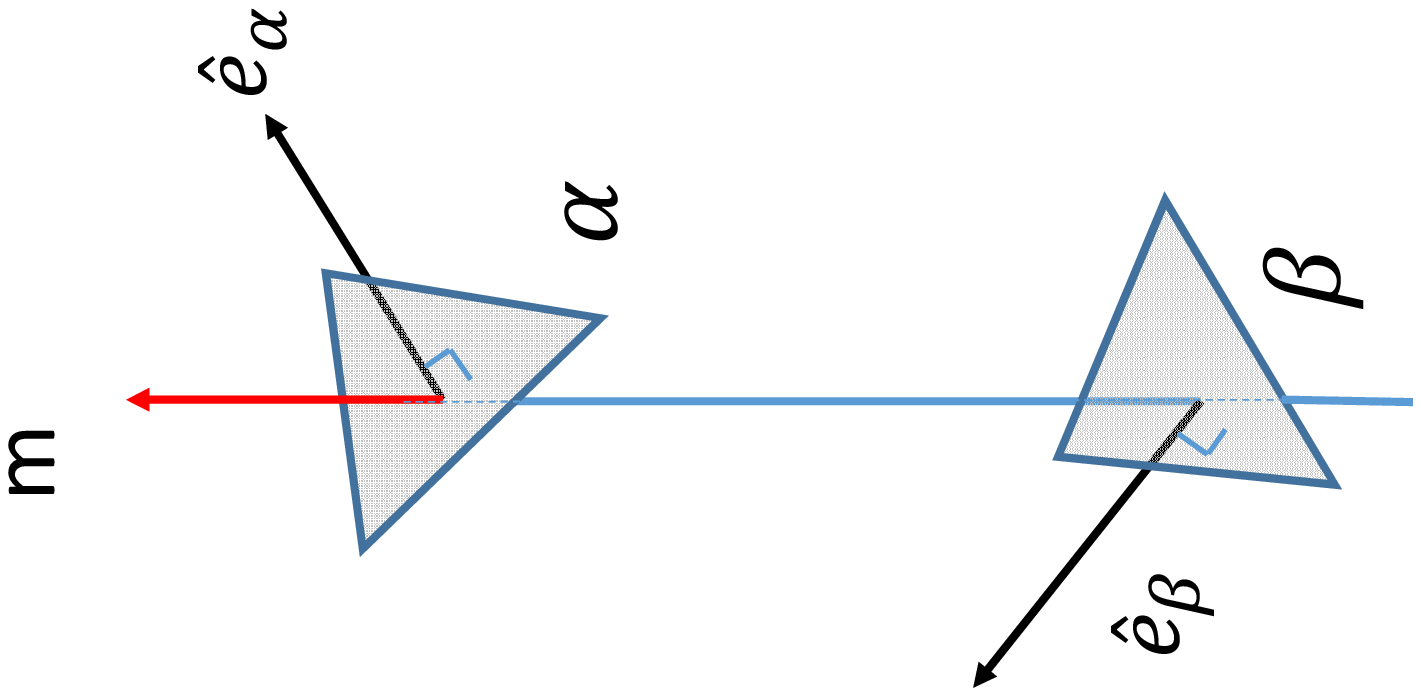}
 \caption{Configuration of the two triangles $\alpha $ and $\beta$. The unit vector $\vem$ shows the direction from $\beta$ to $\alpha$.  We put $\vue_\alpha$ and   $\vue_{\beta}$ as  the unit normal vectors to the triangle planes.   We characterize the geometry of  this system  using the three cosines $\cp\equiv \vue_{\alpha}\cdot \vem,~~\cq\equiv \vue_{\beta}\cdot \vem$ and $\cpq\equiv \vue_{\alpha}\cdot \vue_{\beta}$. 
} 
\end{figure}


 Here the key point is that 
the factors $X_{ij}$ do not depend on the rotations of the tensors $(\veD_A,\veD_E)$  and $(\veD_{A\rq{}},\veD_{E\rq{}})$ respectively  on their detector planes. Then, for the factors $X_{ij}$,   the information of the detector tensors should be completely specified by the unit vectors $\vue_{\alpha}$ and $\vue_{\beta}$ respectively normal to the two detector planes (see Fig.4).  As a result, the factors $X_{ij}$ should be written by the three   vectors $\vue_{\alpha}, \vue_{\beta}$  and $\vem$ with totally six degrees of freedom.   But our observational target is an isotropic  background, and the factors $X_{ij}$ should not be changed by the overall three-dimensional rotation of the network (characterized by the three Euler angles).

For the remaining three degrees of freedom, we select the three cosines given by the inner products of the three unit vectors
\beq
\cp\equiv \vue_{\alpha}\cdot \vem,~~\cq\equiv \vue_{\beta}\cdot \vem,~~\cpq\equiv \vue_{\alpha}\cdot \vue_{\beta}\label{cos}.
\eeq
In fact, we can rewrite the tensorial factors $X_{ij}$ in terms of the three cosines as follows
\beqa
X_{00}&=&\frac1{16}\lmk1+6\cpq^2+\cpq^4  \rmk\label{x1}\\
X_{11}&=&\frac1{16}\lmk1+\cpq^2  \rmk   (1-\cp^2)(1-\cq^2)\\
X_{22}&=&\frac1{16} (1-\cp^2)^2(1-\cq^2)^2\\
X_{01}&=&\frac1{16}[ 1-\cp^2-\cq^2-\cpq\cp\cq+\cpq^3\cp\cq \nonumber \\
       & & -\cpq^2(\cp^2+\cq^2-3) ]\\
X_{02}&=&\frac1{16}[  \cpq^2(1+\cp^2)(1+\cq^2) -4\cpq\cp\cq (\cp^2+\cq^2) \nonumber \\
       & & +(2\cp^2+\cq^2-1)(2\cq^2+\cp^2-1) ] \\
X_{12}&=&\frac1{16} (1-\cp^2)(1-\cq^2)\nonumber\\
       &  & \times (1-\cp^2-\cq^2+\cp\cq\cpq) \label{x6}.
\eeqa
We can easily check these expressions using a software such as $Mathematica$.  Therefore, the total response function $Y_{\alpha\beta}$ is formally expressed as
\beq
Y_{\alpha\beta}=~\sum_{i=0}^2\sum_{j=0}^2 b_i(y) b_j(y) X_{ij} (\cp,\cq,\cpq) \label{yab2}
\eeq
with the argument $y=2\pi f d/c$ for the wave factors  $b_i(y)$.

\subsection{some remarks}
Here we make  some remarks about the tensorial factors.  In general, the three cosines in Eqs.(\ref{cos}) do not uniquely determine the three-dimensional homothetic structure of the network.
As in the case of molecular structures of optically active materials, we  cannot distinguish the configuration shown in Fig.4 from its mirrored counterpart  (with some exceptions including those in Sec. IV).  
 But the two (original and mirrored) networks have an identical sensitivity at least for the parity even quantities as studied in this paper. 

 As shown in Eqs.(\ref{x1})-(\ref{x6}),  the  tensorial factors are given as polynomials of three cosines.  There are three simple relations for the indexes $n_\alpha,n_\beta$ and $n_{\alpha\beta}$ of the involved terms $\cp^{\it n_\alpha}  \cq^{\it n_\beta}\cpq^{\it  n_{\alpha\beta}} $.   
First, if we interchange the two labels $\alpha$ and $\beta$, the unit vector flips $\vem\to -\vem$. Correspondingly, $\cp$ and $\cq$ change their signs, and the summation $n_\alpha+n_\beta$ must be an even number to keep the factors $X_{ij}$ unchanged.  In addition, the factors  should be invariant for the replacement $\vue_{\alpha}\to -\vue_{\alpha}$, and  $n_\alpha+n_{\alpha\beta}$  becomes an even number.   Similarly, $n_\beta+n_{\alpha\beta}$ must be even  for  the flip $\vue_{\beta}\to -\vue_{\beta}$.
We can observe these relations for the power indexes  in  Eqs.(\ref{x1})-(\ref{x6}).

In fact, 
without loss of generality,  we only need to consider the ranges  $\cp\ge 0 $ and $\cq\ge 0$. But, for simplicity, we do not introduce these limitations.

\section{special cases}
Our expression (\ref{yab2}) (together with Eqs.(\ref{b0})-(\ref{b2}) and (\ref{x1})-(\ref{x6})) is given for two triangular detectors with general network configuration as  shown in Fig.4. We can simplify it, when a network has  a special geometrical symmetry. In this section, we briefly discuss three representative examples.  They have inclusion relations.

\subsection{linearly dependent}

If the 
three vectors $ \vue_{\alpha},  \vue_{\beta}$ and $\vem$ are linearly dependent, we can express the three cosines with two parameters  as 
\beq
\cp=\sin\rho , \cq=\sin\sigma,\cpq=\cos(\rho+\sigma).
\eeq
Then we can show 
\beq
X_{11}=X_{02}.\label{1102}
\eeq 


If the two triangles are tangent to a sphere (as for the LISA-Taiji network \cite{Seto:2020zxw,1821247}), the three vectors are linearly dependent. In this case, we can put
\beq
 \cp=\sin(\rho/2), \cq=\sin(\rho/2),\cpq=\cos\rho.
\eeq
Here $\rho$ is the opening angle between the two triangles measured from the center of the sphere. 
After some calculations, we can recover the results presented in the literature \cite{Seto:2020zxw,1821247}
\beq
Y_{\alpha\beta}=2\Theta_1^2+2\Theta_2^2
\eeq
with
\beqa
\Theta_1&=&\cos^4\lmk  \frac{\rho}2\rmk \lmk j_0+\frac57j_2+\frac3{112}j_4  \rmk\\
\Theta_2&=&\lmk-\frac38 j_0+\frac{45}{56}j_2-\frac{169}{896}j_4  \rmk\nonumber\\
& &
+\lmk\frac12 j_0-\frac{5}{7}j_2-\frac{27}{224}j_4  \rmk\cos\rho\nonumber\\
& &+ \lmk-\frac18 j_0-\frac{5}{56}j_2-\frac{3}{896}j_4  \rmk\cos2\rho.
\eeqa
{These expressions might be used for studying correlation analysis with two ET-like triangular detectors on the Earth.}

\subsection{normal direction }
As a subset of linearly dependent configurations, we consider the case when one of the triangle (say $\alpha$) is normal to the the direction vector $\vem$ ($\cp=\pm 1$).  From the relations 
 $D_{A,ij} m_i=D_{E,ij} m_i=0$,    only the factor  $X_{00}=0$ is non-vanishing, and we obtain
\beq
Y_{\alpha\beta}=b_0(y)^2X_{00}(\cpq).
\eeq

\subsection{collocated triangles}
If the barycenter of the two triangles are at the same place ($d=0$), we have $y=0$.  This can be regarded as a subset of the previous subsection. Using the relation
\beq
j_n(0)=\delta_{n0},
\eeq
we obtain
\beq
Y_{\alpha\beta}=4X_{00}(\cpq). \label{qq}
\eeq
Later, we will apply this expression for correlation analysis with two  TianQin units.
Eq.(\ref{qq}) is also the result at the low frequency limit $y\ll 1$ for general network configurations. 

\section{Tensorial factors for LISA, TianQin and Taiji}

In this section, we first present basic geometrical information of LISA, Taiji and TianQin.  Then, we  evaluate the tensorial factors $X_{ij}$ for their representative pairs.

\begin{figure}
 \includegraphics[width=\linewidth,angle=270,clip]{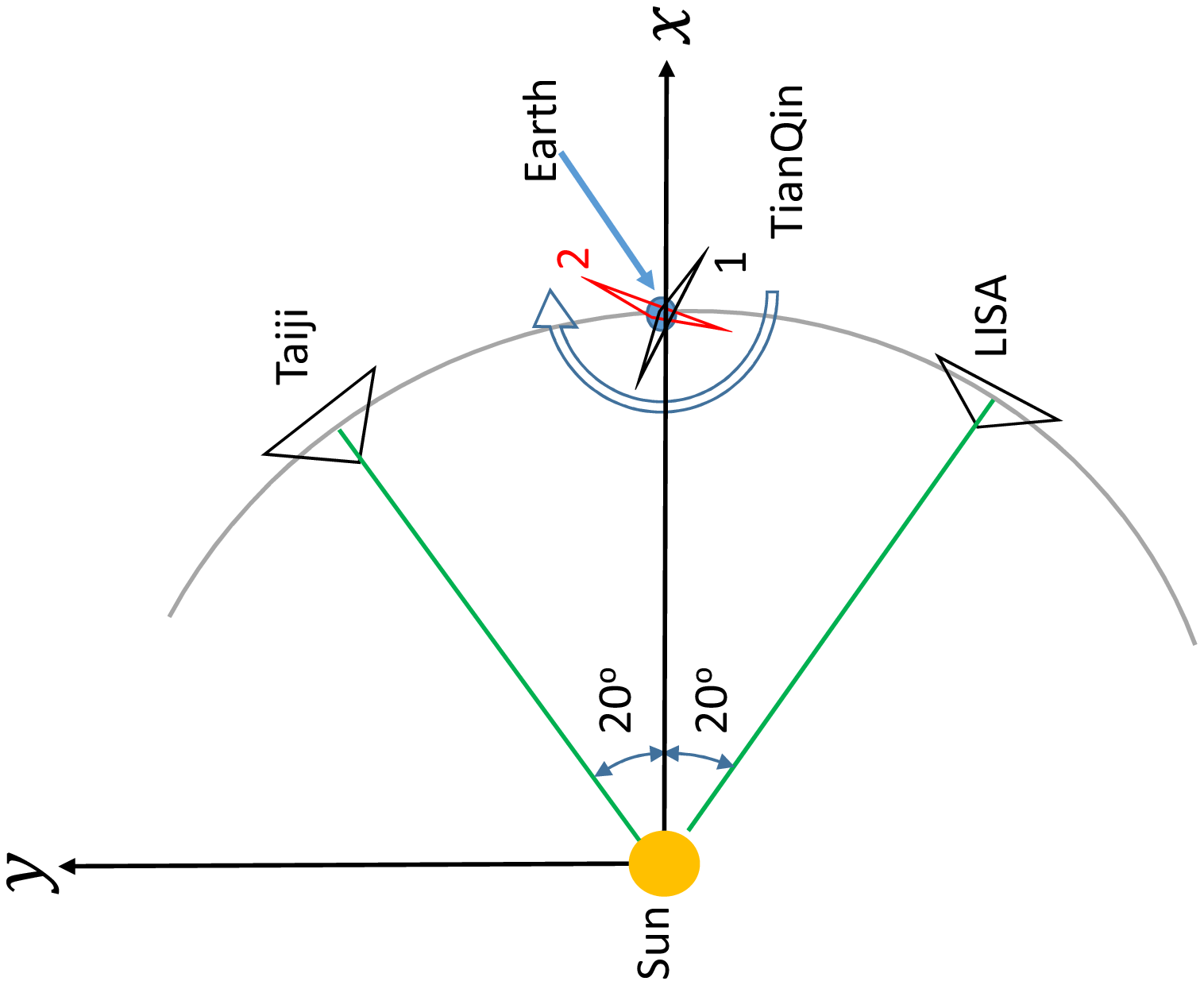}
 \caption{Schematic picture for the configurations of LISA, Taiji and TianQin. We use the frame co-rotating with the Earth.  The $x
$- and $y$-axes are on the ecliptic plane. 
The two triangles for LISA and Taiji are effectively fixed in this frame. TianQin is a geocentric project and uses up to 2 triangle units; TianQin1 and TianQin2.  Their normal vectors rotate in 1 year period.  
} 
\end{figure}


\subsection{basic geometrical parameters}
Fig.5 shows the planned positions of the three projects in the coordinate system co-rotating with the Earth around the Sun.   Here the $xy$-plane is the ecliptic plane.
As for the orientation of each triangles, considering the argument in section III E, we just need to specify their normal vectors. 

LISA is a heliocentric detector at $20^\circ$ behind the Earth. In the co-rotating frame, its normal vector is given by 
\beq
\vue_{L}=\lmk  -\frac{\sqrt{3}\cos20^\circ}2 ,  \frac{\sqrt{3}\sin20^\circ}2,\frac12 \rmk
\eeq
with the label $L$ for LISA.

The configuration of Taiji  (labeled $J$) is similar to LISA but at $20^\circ$ ahead of the Earth. 
Its normal vector is given by 
\beq
\vue_{J}=\lmk  -\frac{\sqrt{3}\cos20^\circ}2 ,  -\frac{\sqrt{3}\sin20^\circ}2,\frac12 \rmk.
\eeq
For comparison, we consider a mirrored image of the Taiji triangle with respect to the ecliptic plane \cite{1821247}. The resultant triangle (labeled $J\rq{}$) can be composed by three heliocentric orbits,  and its normal vector is given by 
\beq
\vue_{J\rq{}}=\lmk  -\frac{\sqrt{3}\cos20^\circ}2 ,  -\frac{\sqrt{3}\sin20^\circ}2,-\frac12 \rmk.
\eeq
For the $LJ$ pair, we have a virtual sphere that  simultaneously contact with  the two triangles \cite{Seto:2020zxw,1821247}. But this is not the case for the $LJ\rq{}$ pair.

TianQin is a geocentric project and  is planned to use  up to two triangle units; TianQin1 and 2 (respectively with the labels $Q$ and $Q\rq{}$) \cite{Huang:2020rjf}.  The normal vector of  TianQin1 is given by
\beq
\vue_{Q}(t)= (-\cos\phi\cos\zeta,\sin\phi\cos\zeta,-\sin\zeta) \label{vq}
\eeq
that rotates at 1 year period parameterized by 
\beq
\phi\equiv 2\pi \lmk \frac{t}{\rm 1yr} \rmk.  \label{phi}
\eeq
In a non-rotating frame, the normal vector $\vue_{Q}$ is fixed and directed to the known binary 
 RX J0806.3+1527 \cite{Luo:2015ght,Huang:2020rjf}.  Here $\zeta=4.7^\circ$  is the small offset angle between the binary and the ecliptic plane.

The normal vector of TianQin2 (in the co-rotating frame) is given by 
\beq
\vue_{Q\rq{}}(t)= [-\cos(\phi-\pi/2),\sin(\phi-\pi/2),0]
\eeq
with the same parametrization as Eq.(\ref{phi}) \cite{Huang:2020rjf}. We have $\vue_{Q\rq{}}(t)\cdot \vue_{Q}(t)=0$. 
 Even with the small misalignment angle    $\zeta$,  we also have 
$\vue_{Q\rq{}}(t+1/4)\simeq \vue_Q(t)$.

In the original report of TianQin \cite{Luo:2015ght},  the basic mission parameters are presented for a single triangular unit whose normal vector is identical to $\vue_Q$.  According to the report, the observational windows of the unit might be $2\times(3\,\rm months)$ per one year.   This reflects the instrumental design to suppress the mission cost in relation to the solar radiation. 
If we simply follow this operation schedule, taking into account of the relative direction to the Sun, the  duty cycles of TianQin1 and TianQin2 would  be  at most 50\%,  without overlapped period. 
In this case, the possibility of correlation analysis is excluded. Below, we optimistically assume that the two units have 100\% duty cycles (thus completely overlapped, see e.g. \cite{Huang:2020rjf}) 

Using the geometrical parameters presented in this subsection, we next evaluate the tensorial factors $X_{ij}$ and the  total response functions $Y_{\alpha\beta}(f)$ for some of the detector pairs. 

\subsection{LISA-Taiji network}

The LISA-Taiji network has the separation 
$d=1{\rm AU}\times 2\sin20^\circ=1.0\times 10^8$km with the unit directional vector
\beq
\vem_{}=(0,1,0)~~
\eeq
 in the co-rotating frame.
Then, using expressions (\ref{x1})-(\ref{x6}), we have 
\beqa
X_{00}&=&0.346,X_{11}=0.087,X_{22}=0.043,\\
X_{01}&=&0.173,X_{02}=0.087,X_{12}=0.039.
\eeqa
As shown in Eq.(\ref{1102}), we have $X_{11}=X_{02}$ for this symmetric network.  We present the  total response function $Y_{LJ}$ in Fig.6. We have $Y_{LJ}(f)\ll 1 $ around $f\sim 2$mHz, as pointed out in \cite{Seto:2020zxw,1821247}.   


We also evaluate the tensorial factors for the $LJ\rq{}$ pair and obtain
\beqa
{X}_{00}&=&0.103, {X}_{11}=0.057,{X}_{22}=0.043, \\
{X}_{01}&=&0.072, {X}_{02}=0.043, {X}_{12}=0.041.
\eeqa
As shown in Fig.6,  the asymptotic value $Y_{LJ\rq{}}(f=0)$ is smaller than that of the $LJ$ pair, but we have $Y_{LJ\rq{}}(f)\gg Y_{LJ}(f)$ around 2mHz.

\begin{figure}
 \includegraphics[width=8.cm,angle=0]{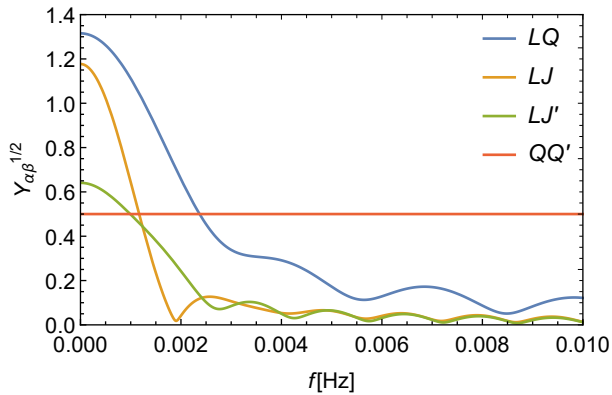}
 \caption{The  total response function $Y_{\alpha\beta}(f)$ for the four pairs; LISA-TianQin ($LQ$), LISA-Taiji ($LJ$), its variation ($LJ\rq{}$) and  TianQin1-TianQin2 ($QQ\rq{}$).   
}
\end{figure}


\subsection{LISA-TianQin network}

The LISA-TianQin1 ($LQ$) network has the distance 
$d=1{\rm AU}\times 2\sin10^\circ=5.2\times 10^7$km and the unit directional vector 
\beq
\vem=(-\cos80^\circ,-\sin20^\circ,0).~~
\eeq
While the wave factors $b_i(y)$ are time independent,  the tensorial factors $X_{ij}$ change with time due to the annual rotation of the normal vector $\vue_{Q}$ as in Eq.(\ref{vq}).  But we can easily take their time averages appropriate for the total signal-to-noise ratio; 
\beq
\bar{X}_{ij}=\frac1{2\pi}   \int_{0}^{2\pi} X_{ij}d\phi. \label{ti}
\eeq

 We numerically obtain 
\beqa
\bar{X}_{00}&=&0.432, \bar{X}_{11}=0.095, \bar{X}_{22}=0.045,  \label{lq1}\\
\bar{X}_{01}&=&0.186, \bar{X}_{02}=0.086, \bar{X}_{12}=0.045,\label{lq2}
\eeqa
and plot the  total response function $Y_{LQ}(f)$ in Fig.6 with
\beq
{Y}_{LQ}=~\sum_{i=0}^2\sum_{j=0}^2 b_i(y) b_j(y) {\bar X}_{ij}.
\eeq

Similarly we can evaluate the tensorial factors for  the LISA-TianQin2 ($LQ\rq{}$) network. The two networks $LQ$ and $LQ\rq{}$ have slightly different tensorial factors due to the small misalignment angle $\zeta$. But,  except for $X_{00}=0.432104$($LQ$)  and    $X_{00}=0.432617$$(LQ\rq{})$, they have no difference in the significant digits used in  Eqs.(\ref{lq1}) and (\ref{lq2}).    From  the geometrical symmetry, we also have $Y_{LQ}= Y_{JQ}$ and $Y_{LQ\rq{}}= Y_{JQ\rq{}}$.

\subsection{TianQin1-2 network}
 From Eqs.(\ref{qq}) with $\cpq=\vue_Q\cdot \vue_{Q\rq{}}=0$, we have
\beq
Y_{QQ\rq{}}=\frac14.
\eeq

\section{signal-to-noise ratio}

As shown in Eq.(\ref{snr0}), for two triangle detectors $\alpha$ and $\beta$, the total signal-to-noise ratio is given by 
\beq
SNR_{\alpha\beta}^2=\lmk \frac{3H_0^2}{10\pi^2} \rmk^2 T_{\rm obs}  \lkk 2\int_{f_{\rm min}}^{f_{\rm max}}  df \frac{\Omega_{\rm GW}(f)^2 Y_{\alpha\beta}(f)}{f^6N_{\alpha}(f)N_{\beta}(f)}   \rkk  \label{snb}
\eeq
with the  total response function $Y_{\alpha\beta}(f)$.  In this section, we first present the noise spectra $N_\alpha(f)$ for the three projects and then numerically evaluate the total signal-to-noise ratios.  Note that, even  for the rather complicated LISA-TianQin1 pair, with our formulation based on Eq.(\ref{yab2}), we can separately deal with the time and frequency integrals (see also Eq.(\ref{ti})).

\subsection{instrumental noise spectra }

The arm length of LISA is 
$R_L=2.5\times 10^6$km  with the  associated characteristic frequency $f_L=c/(2\pi R_L)=0.019$Hz.  The low frequency approximation is efficient in the range $f<f_L$.  The noise spectrum of LISA\rq{}s $A$ and $E$ channels is approximately given by \cite{Cornish:2018dyw}
\beqa
N_L(f)&=&\frac4{3R_L^2} \lkk  P_{o1} +2 [1+\cos(f/f_L)^2] \frac{P_{a1}}{(2\pi f )^4} \rkk \nonumber \\
& &\times [1+0.6(f/f_L)^2]   \label{nl}.
\eeqa 
Here the acceleration noise and the optical path noise are respectively given by 
\beqa
P_{a1}&=&9.0\times 10^{-30} [1+(4\times 10^{-4}/f)^2]\nonumber\\
 & & \times (1+[f/(8\times 10^{-3})]^4){\rm m^2 s^{-4} Hz^{-1}} \label{pa1}
\eeqa
\beq
P_{o1}=2.25\times 10^{-22} [1+(2\times 10^{-3}/f)^4] {\rm Hz^{-1}}\label{po1}.
\eeq
In Eq.(\ref{nl}) we did not include the angular averages for the beam pattern functions (a factor 2/5 times smaller than \cite{Cornish:2018dyw}). The noise spectrum $N_L(f)$ is presented in Fig.1.

Taiji has the arm length
$R_J=3.0\times 10^6$km  with the characteristic frequency $f_J=c/(2\pi R_J)=0.016$Hz.  Using the same functional forms as Eqs.(\ref{pa1}) and (\ref{po1}), its noise  spectrum is given by \cite{Wang:2020vkg}
\beqa
N_J(f)&=&\frac4{3R_b^2} \lkk 0.8^2 P_{o1} +2 [1+\cos(f/f_b)^2] \frac{P_{a1}}{(2\pi f )^4} \rkk \nonumber \\
& &\times [1+0.6(f/f_b)^2] .
\eeqa 
We put $ N_{J\rq{}}(f)=N_J(f)$ for the slightly different configuration.

The triangular units of TianQin have the arm length  
$R_Q=1.7\times 10^6$km much smaller than LISA and Taiji.  The characteristic frequency for the low frequency approximation is 
$f_Q=c/(2\pi R_Q)= 0.28$Hz. 
The noise spectrum for its $A$ and $E$ data channels is approximately given by \cite{Huang:2020rjf}
\beqa
N_Q(f)&=&\frac4{3L_Q^2}   \lkk P_{o2}+  \frac{4P_{a2}}{(2\pi f )^4}  \lmk 1+\frac{\rm 0.1mHz}f \rmk  \rkk \nonumber \\
& &\times [1+0.6(f/f_Q)^2] 
\eeqa 
with 
$P_{a2}=1.0\times 10^{-30} {\rm m^2 s^{-4} Hz^{-1} }      $ and   $P_{o2}=1.0\times 10^{-24} {\rm m^2 Hz^{-1}}$.

\subsection{numerical results }

\begin{figure}
 \includegraphics[width=8.cm,angle=0]{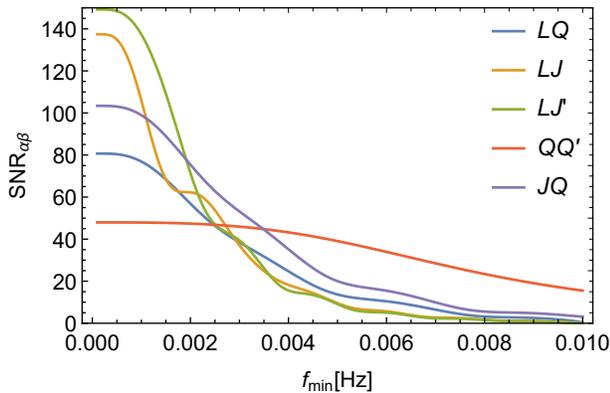}
 \caption{The signal-to-noise ratios $SNR_{\alpha\beta}$  as functions of the low frequency cut-off frequency $f_{\rm min}$. We assume a flat  spectrum  $\Omega_{GW}=10^{-11}$ and the integration period $T_{obs}=10$yr. 
} 
\end{figure}


Now we numerically evaluate the total $SNR_{\alpha\beta}$ for some pairs of detectors.  As a fiducial model of the background spectrum $\Omega_{GW}(f)$, we use the following flat model 
\beq
\Omega_{GW}=10^{-11}.
\eeq
In Eq.(\ref{snb}), we take the maximum frequency at $f_{\rm max}=0.1$Hz and fix the integration period at $T_{\rm obs}=10$yr. We set the minimum frequency $f_{\rm min}$ as a free parameter. In reality, the frequency $f_{\rm min}$ should be determined by the subtraction of the Galactic binary foreground, and is closely related to the integration period $T_{\rm obs}$ \cite{Cornish:2018dyw}.  In  typical astronomical models, the expected value would be $f_{\rm min}= 2\sim$5mHz for  observational periods 4-10yr.

In Fig.7, our numerical results  are presented as functions of $f_{\rm min}$. Using this figure, we first discuss the validity of the low frequency approximation for our calculations. As mentioned in the previous subsection, this approximation does not work above the characteristic frequency $f_\alpha=c/(2\pi R_\alpha)(\ge f_J=16$mHz) determined by the arm length $R_\alpha$ of the detector $\alpha$.  But as shown in Fig.7,  the contributions of the signals above 8mHz are negligible, except for the $QQ\rq{}$-pair.  Even in the case of the $QQ\rq{}$-pair, we have only $SNR_{QQ\rq{}}=1.12$ for $f_{\rm min}=30$mHz$\ll f_Q=$280mHz, and the contribution from $f\gsim f_Q$ is negligible.    Therefore, in all cases, we can safely apply the low frequency approximation for estimating $SNR_{\alpha\beta}$ presented in Fig.7. The signal-to-noise ratio is given by the frequency integral (\ref{snb}) of the product $Y_{\alpha\beta}(f)\cdot f^{-6}\cdot (N_\alpha N_\beta)^{-1}$ and the factor $f^{-6}$ strongly suppresses the contribution of the high frequency part. Unless the input spectrum $\Omega_{GW}(f)$ is heavily blue tilted,  the low frequency approximation works efficiently.

{For the combination of a realistic cut-off $f_{\rm min}=2.5$mHz and a somewhat optimistic overlapped time $T_{\rm obs}=10$yr, the signal-to-noise ratios are in a relatively narrow range 50-75.  For the threshold $SNR_{\alpha\beta}=10$, the detection  limits are roughly given by $\Omega_{GW}\sim 2\times 10^{-12}$. 
}

Next, we discuss the characteristic aspects of individual curves in Fig.7.   The result for the $LJ$ pair was already studied in \cite{Seto:2020zxw}, but we briefly mention its basic profile for comparison with other cases.   In Fig.7, for the $LJ$ pair, we can observe a flat region around $f_{\rm min}\sim2$mHz. This reflects the deep dip of the  total response function $Y_{LJ}\ll 1$ at $f\sim2$mHz (see Fig.6).  For the $LJ\rq{}$-pair, we can observe a similar but less prominent modulations around 3mHz,  4mHz and 6mHz, corresponding  to the dips of the  total response function $Y_{LJ\rq{}}$.  On the whole, affected by the difference around 2mHz, we have $SNR_{LJ}<SNR_{LJ\rq{}}$ for $f_{\rm min}\lsim 2.3$mHz.  
This inequality  might be interesting for planning potential collaboration between LISA and Taiji. 

As shown in Fig.7, the $QQ\rq{}$-pair could be a powerful probe for the frequency regime $f\gsim$5mHz, compared with other combinations.  This curve is for our optimistic assumption with 100\% duty cycle in 10 years.   Even with the total overlapped time of 2 years, the expected signal-to-nose ratio is obtained by multiplying the factor $0.2^{1/2}=0.45$ to the result $SNR_{QQ\rq{}}$ given in Fig.7.  We can still keep the sensitivity level $\Omega_{GW}\lsim 10^{-11}$.  

In Fig.7, except for the overall scaling, the two curves for $SNR_{LQ}$ and $SNR_{JQ}$ are quite similar. Due to the symmetry of the networks $LQ$ and $JQ$, we have $Y_{LQ}=Y_{JQ}$ and the difference between the two curves is determined by the noise spectra $N_L(f)$ and $N_T(f)$ (especially at $f\lsim6$mHz).

\section{summary and discussion}

In this paper, we discussed the detectability of an isotropic gravitational wave background by correlating multiple triangular  detectors such as LISA, TianQin and Taiji. In general network configurations, we still have the rotational symmetry of the data channels (Sec.III D) and, consequently, the tensorial factors $X_{ij}$ are written by  the inner products of three unit vectors $\vem$, $\vue_\alpha$ and $\vue_\beta$ (see Fig.4).  With our expression (\ref{yab2}), we can evaluate the total response function $Y_{\alpha\beta}$ without directly dealing with detector tensors.

We evaluated the expected signal-to-noise ratios for various pairs of detectors, including correlation between LISA ($L$) and the slightly different version ($J\rq{}$) of Taiji ($J$). For a cut-off frequency  $f_{\rm min}\lsim 2.3$mHz, we have $SNR_{LJ\rq{}}>SNR_{LJ}$, strongly affected by the dip $Y_{LJ}\ll 1$ around 2mHz (see Fig.6). 

We also pointed out that, as shown in Fig.7, the two TianQin units can play unique role to probe a background at the relatively high frequency regime $\gsim$5mHz.  But we need to simultaneously operate its two units.  It would be interesting to consider such feasibility for pursuing the advantage of the TianQin project.

In this paper, we focused our attention on the isotropic intensity of a gravitational wave background, as the most fundamental quantity. To study the background in full detail, we should examine additional properties such as the polarization states. 
The separation between detectors are essential for probing some of these properties \cite{Seto:2020zxw,1821247,Nishizawa:2009jh}. 
 As demonstrated in this paper, we will be able to develop simplified formulations also for the additional properties, using the three vectors characterizing the network geometry.

We have assumed that the correlation of instrumental noises is ignorable  for two triangular detectors. But the actual correlation level should be carefully examined by using real data streams. Such studies can be  important for the potential follow-on missions such as DECIGO \cite{Seto:2001qf,Kawamura:2011zz} and BBO \cite{Harry:2006fi}.

\begin{acknowledgments}
The author would like to thank H. Omiya for useful conversations. 
 This work is supported by JSPS Kakenhi Grant-in-Aid for Scientific Research
 (Nos. 17H06358 and 19K03870).
\end{acknowledgments}

\end{document}